# SUMMARY OF WORKING GROUP G: BEAM MATERIAL INTERACTION*†

D. Kiselev[#], Paul Scherrer Institute, Switzerland
N.V. Mokhov, Fermilab, Batavia, IL 60510, U.S.A.
R. Schmidt, CERN, Geneva, Switzerland

## Abstract

For the first time, the workshop on High-Intensity and High-Brightness Hadron Beams (HB2010), held at Morschach, Switzerland and organized by the Paul Scherrer Institute, included a Working group dealing with the interaction between beam and material. Due to the high power beams of existing and future facilities, this topic is already of great relevance for such machines and is expected to become even more important in the future. While more specialized workshops related to topics of radiation damage, activation or thermo-mechanical calculations, already exist, HB2010 provided the occasion to discuss the interplay of these topics, focusing on components like targets, beam dumps and collimators, whose reliability are crucial for a user facility. In addition, a broader community of people working on a variety of issues related to the operation of accelerators could be informed and their interest sparked.

<>
[*]Work supported by Fermi Research Alliance, LLC under contract No. DE-AC02-07CH11359 with the U.S. Department of Energy.
[†]Presented paper at 46th ICFA Advanced Beam Dynamics Workshop on High-Intensity and High-Brightness Hadron Beams, Sept. 27 - Oct. 1, 2010, Morschach, Switzerland.
[#]Daniela.Kiselev@psi.ch

# SUMMARY OF WORKING GROUP G: BEAM MATERIAL INTERACTION

D. Kiselev[#], Paul Scherrer Institute, Switzerland
N.V. Mokhov, Fermilab, Batavia, IL 60510, U.S.A.
R. Schmidt, CERN, Geneva, Switzerland

*Abstract*
For the first time, the workshop on High-Intensity and High-Brightness Hadron Beams (HB2010), held at Morschach, Switzerland and organized by the Paul Scherrer Institute, included a Working group dealing with the interaction between beam and material. Due to the high power beams of existing and future facilities, this topic is already of great relevance for such machines and is expected to become even more important in the future. While more specialized workshops related to topics of radiation damage, activation or thermo-mechanical calculations, already exist, HB2010 provided the occasion to discuss the interplay of these topics, focusing on components like targets, beam dumps and collimators, whose reliability are crucial for a user facility. In addition, a broader community of people working on a variety of issues related to the operation of accelerators could be informed and their interest sparked.

## TALK AND POSTER SESSIONS

Even though organized for the first time, working group G had three sessions with 15 talks, additional two talks in a joint session with working group A "Beam Dynamics in High-Intensity Circular Machines" as well as five posters. Various topics were covered:

- Activation: nuclide inventory and dose rates
- Radiation damage: calculations and experiments
- Thermo-mechanical simulations: design tools for targets, collimators and beam dumps
- Irradiation facilities: existing and upcoming
- Future accelerator facilities: upgrades and plans

The last topic in the list, the upgrade of existing and plans of new facilities to operate with even more beam power, drives the need to seriously address the other topics. While one path for future accelerators is to increase the beam energy, the other is to increase the beam current. The product of both is an increase of the beam power which finally has to be deposited somewhere - usually on targets, collimators and beam dumps.

[#]Daniela.Kiselev@psi.ch

These components get highly activated and their nuclide inventory has to be determined, when they finally get disposed as radioactive waste. In addition, for maintenance, dose rates have to be known in advance to plan working procedures and as design criteria in the development phase of new components (as it was done for the LHC beam dumps, S. Roesler, CERN). For these purposes, particle transport Monte Carlo codes like FLUKA (S. Roesler, CERN) and MARS15 (N. Mokhov, Fermilab) are employed. Improvements in the predictive power of the codes were made and benchmarks with experimental data were performed. Recently their capabilities were significantly extended and new features added. These activities were driven on one hand by user demands, on the other hand by applications, e.g. for the LHC. The calculation and use of $H_2$ and He gas production and of the quantity "Displacements Per Atom (DPA)", a measure of radiation damage, is an attempt to compare damage caused by radiation under different conditions. Regarding DPA, there are still discrepancies between different codes, which have to be solved in the near future. Another issue related to the activation of components is the growing interest in choosing materials which get less activated but have the same or equivalent mechanical and physical properties needed for the application (J.H. Jang, KAERI, E. Mustafin, I. Strasik, GSI).

Due to the high power deposition which is dissipated as heat, efficient cooling systems have to be designed. Tensile stress induced by thermal expansion has to be kept within the safety margin. For this purpose thermo-mechanical simulations are performed using commercial tools like ANSYS and CFD-ACE. Thermal, mechanical and electromagnetic models can be coupled and applied to a detailed geometry. As input, the energy deposition due to the particle beam are taken from particle transport Monte Carlo codes or from a subroutine implemented into the multiphysics program as done for CFD-ACE by Y.J. Lee (PSI). Examples of components suffering from heavy power load are the T2K target at JPARC with 750 kW (J. Densham, STFC/RAL), 200 kW on a Cu collimator at PSI (Y. Lee) and the Neutrino beam factory at Fermilab, which plans to start with 700 kW power load and upgrade later to up to 2 MW (P. Hurh, Fermilab). Thermal and stress simulations for Conceptual Design Studies are underway for two target alternatives made of beryllium and graphite, respectively. At FRIB an extreme high power density of 20 to 60 MW/cm$^3$ for the pulsed ion beam is expected (R. Ronnigen, NSCL-FRIB). To predict the behavior of components under irradiation, it would be important to include the change of mechanical and physical properties. The problem is that these parameters are often not known for the required conditions. In addition, more precise lifetime predictions are needed. At the moment, one is obliged to use conservative limits due to the large uncertainties included. E.g., it is expected that the FRIB target will last only for two weeks. Another effect observed due to the irradiation with high-energy protons is the decrease of the neutrino production rate from the graphite target used at Fermilab (S. Striganov, Fermilab). This might be due to atom displacement or the production of helium, which would lead to a density reduction. Examinations of the target material are underway. It is interesting that MARS15 calculations revealed that the relative distribution of DPA is very similar to the amount of helium produced. This might offer a possibility for an indirect measurement of DPA by determining the He-content.

Whereas a lot of material data under thermal neutron irradiation is available, not much exists for high-energy particles. To profit from the existing database, it is required to relate the damage caused by low-energy particles to the one made by high-energy particles. This is called damage correlation (M. Li, ANL). It is a very complex problem, because the change of material properties depends on many other conditions like the temperature and the DPA rate. To perform this task, dedicated irradiation test experiments using high-energy particles under different conditions, i.e. varying parameters like the energy, temperature, irradiation times and particle types are needed. The long-term perspective is to predict the change of material properties by (phenomenological) models. A prerequisite is a reliable prediction of DPA and the gas production rates (He, H).

N. Simos (BNL) reported on experiments done at the BLIP irradiation station at BNL. Different materials, including new generation materials and composites, were irradiated with 200 MeV protons to study the change of mechanical and physical properties under varying conditions in a systematic way to shed some light on damage correlations. A new irradiation facility, HiRadMat at CERN will go into operation at the end of 2011 (presented by R. Losito (CERN)). The facility is designed to test shock impact of beam on materials, about $10^{16}$ protons at 450 GeV will be available per year. The demand is driven by the machine protection of the present and future LHC. Aside from the examination of material properties under these conditions, fields of investigations will be the failure from pulsed beam impact, shock wave generation and propagation and the validation of the tunneling effect (J. Blanco, CERN and N.A. Tahir, GSI). For the new facility FAIR, T. Seidl (TU Darmstadt) investigated insulation materials for the superconducting SIS-100 dipole magnets under very different irradiation conditions. Ions, protons and neutrons ranging from a few MeV to 800 MeV as well as photons from a $^{60}$Co source were used. Thermal properties like the thermal conductivity and the specific heat as well as the breakdown voltage were measured at low temperatures.

## DISCUSSION SESSION

The community agreed that irradiation experiments with high-energy particles are very important and urgently needed. Models require benchmarking from experimental data to make predictions for the change of material properties under irradiation in the future. Since the nuclear reaction cross section stays almost constant for energies above 200 MeV, energies of about 1 GeV or less are sufficient for this purpose. A list of specifications of existing or upcoming irradiation facilities at CERN, Fermilab, GSI, PSI, BNL, Los Alamos, Kurchatov Institute as well as a list of possible users should be made as soon as possible. One problem is that some laboratories have the possibility to irradiate samples but not the infrastructure to test and handle the radioactive material in the hotcell using remote controlled devices. Transportation of radioactive material to another institute is possible but, depending on the activity and distance to be traveled (crossing borders), it requires a lot of administrative effort and money to fulfill the regulations.

Questions were addressed how the handbooks of Los Alamos and ITER, which include a lot of materials data under irradiation, can be made available to a broader community, which is not working in the USA and is not part of the ITER collaboration. Further, the idea came up to establish a handbook for materials irradiated and tested in accelerators. This would be very useful but requires a lot of work and effort. For the moment, nobody volunteered for this job.

## OUTLOOK

The vivid discussions after the talks and in the discussion session itself showed the interest and enthusiasm of the community. It was suggested to extend the High-Power Targetry Workshop, which will take place in Lund 2-6 May 2011, to discuss topics regarding radiation damage, in particular to establish a joint effort for an experimental irradiation program. This includes also a wish list of specifications and where to perform these experiments.